# Evidence and implications of abnormal predictive coding in dementia

Kocagoncu, E.[1,3], Klimovich-Gray, A.[2], Hughes, L.[1,3] & Rowe, J. B.[1,3]

1. Cambridge Centre for Frontotemporal Dementia, Department of Clinical Neurosciences, University of Cambridge, Cambridge, UK
2. Basque Center on Cognition, Brain and Language, San Sebastian, Spain
3. Medical Research Council Cognition and Brain Sciences Unit, University of Cambridge, Cambridge, UK

Correspondence to: Ece Kocagoncu
Cambridge Centre for Frontotemporal Dementia and Related Disorders,
University of Cambridge, MRC Cognition and Brain Sciences Unit
15 Chaucer Road,
Cambridge, CB2 7EF, UK
Email: ek390@cam.ac.uk
Phone: 01223 769879

Number of pages: 16
Number of display items: 3
Numbers of tables: 0
Number of words for abstract: 193
Number of words for introduction: 404
Number of words for conclusion: 216

# Abstract

The diversity of cognitive deficits and neuropathological processes associated with dementias has encouraged divergence in pathophysiological explanations of disease. Here, we review an alternative framework that emphasises convergent critical features of pathophysiology, rather than the loss of "memory centres" or "language centres", or singular neurotransmitter systems. Cognitive deficits are interpreted in the light of advances in normative accounts of brain function, based on predictive coding in hierarchical neural networks. The predicting coding rests on Bayesian integration of beliefs and sensory evidence, with hierarchical predictions and prediction errors, for memory, perception, speech and behaviour. We describe how analogous impairments in predictive coding in parallel neurocognitive systems can generate diverse clinical phenomena, in neurodegenerative dementias. The review presents evidence from behavioural and neurophysiological studies of perception, language, memory and decision-making. The re-formulation of cognitive deficits in dementia in terms of predictive coding has several advantages. It brings diverse clinical phenomena into a common framework, such as linking cognitive and movement disorders; and it makes specific predictions on cognitive physiology that support translational and experimental medicine studies. The insights into complex human cognitive disorders from the predictive coding model may therefore also inform future therapeutic strategies.

# Abbreviations

nvPPA: non-fluent primary progressive aphasia; PSP: progressive supranuclear palsy

# Introduction

Cognitive deficits in neurodegenerative diseases have often been characterised as the loss of core functional modules in distinct brain regions, such as "memory centres" or "executive centres". This approach emphasises the functional difference between disorders, at a time when preclinical models suggest convergence in the pathophysiology of different diseases. Here we re-evaluate diverse cognitive and behavioural features of dementia in terms of advances in predictive coding accounts of brain function (Rao and Ballard, 1999; Friston, 2005a; Bar, 2007; Clark, 2013). We re-assess clinical deficits in terms of the disruptions in a precisely tuned hierarchy of prediction, prediction error and inference.

The predictive coding accounts of normative brain function integrate cognitive and computational neuroscience to explain how we perceive and interact with our environment. It proposes that in health, the brain acts as an active Bayesian inference machine that learns in terms of statistical regularities in the external world (**Box 1**), and generates predictions to increase the efficiency of information processing and understanding of our sensorium (Rao and Ballard, 1999; Friston, 2005a; Bar, 2007; Clark, 2013).

The predictive coding account serves as a common neurobiological framework to describe cognitive, perceptual and behavioural phenomena. There is direct evidence for predictive coding of vision (Hosoya *et al.*, 2005; Hohwy *et al.*, 2008), rhythmic perception (Vuust *et al.*, 2009; Vuust and Witek, 2014), auditory processing (Wicha *et al.*, 2004; Kumar *et al.*, 2011; Dikker and Pylkkänen, 2013; Lewis and Bastiaansen, 2015; Lewis *et al.*, 2015), reward and preferences (O'Doherty *et al.*, 2006), and action control (Ramnani and Miall, 2004; Kilner, 2011). The representation of predictions, prediction errors and precision in each system depends on a fine-tuned cortical hierarchy, with laminar-specific connectivity and balanced excitatory-inhibitory neurochemistry (**Figure 1A**). Imbalances or disruption in the system result in domain-specific or domain-general cognitive impairments, as has been established for psychosis (Fletcher and Frith, 2009; Friston *et al.*, 2014b) and autism (Pellicano and Burr, 2012; Lawson *et al.*, 2014). For example, hallucinations and delusions arise from faulty precision-weighting of the prediction error signals (Fletcher and Frith, 2009), leading to an increased reliance over the internal model and reduced reliance on sensory evidence.

In this Update, we re-assess the impairments in perception, action and higher cognition in the predictive coding framework, and consider the mechanisms of impairment in dementia and related neurodegenerative diseases. We start with perception and action, with an emphasis on the lower levels of cortical hierarchy, before considering higher cognitive systems.

# Perception

In perceiving our environment, we make use of prior knowledge and context to predict sensory inputs. In the auditory scene analysis, we parse its constituent objects over time and space, such as recognising one's own name in a noisy environment (i.e. the cocktail party effect) (Bregman, 1990). Top-down predictions based on prior experience of the speakers, their language and the topic, facilitates this segregation, especially in noisy environments (Griffiths and Warren, 2002). In vision, the context-based predictions likewise aid rapid object recognition under both normal and challenging conditions (Bar, 2007; Summerfield and de Lange, 2014).

The use of auditory predictions is largely preserved in normal ageing (Moran *et al.*, 2013) but can be significantly disrupted in mild cognitive impairment and dementia. These abilities use temporo-parietal areas that are affected by Alzheimer's disease (Golden *et al.*, 2015), and accordingly, patients have difficulty following conversations in the presence of background noise. Patients with Alzheimer's disease show impairments in segregating, tracking and grouping auditory objects that evolve over time (Goll *et al.*, 2012), and in perceiving sound location and motion (Golden *et al.*, 2015). They are

> ## Box 1: Predictive coding and the hierarchical networks
>
> Predictive coding describes how the brain perpetually creates and evaluates its own predictions across cognitive and behavioural domains. To explain this fundamental mechanism of the brain, predictive coding rests their premises on prior cognitive models such as those of Helmholtz, who proposed that perception is the outcome of probabilistic inferences and the predictive dynamics of information processing. The predictive coding accounts (Rao and Ballard, 1999; Friston, 2005a; Bar, 2007; Clark, 2013) put forward a biological implementation of generative models with multi-level hierarchies encompassing neural circuits at the micro and macro level. These models are proposed to capture statistical patterns and dependencies in the external world and events, and used these patterns to deliver top-down predictions, in turn increasing the efficiency of an organism's information processing and understanding of the external world. Each layer in the hierarchy predicts the activity in the layer below through top-down relay of information. When a mismatch arises between the prediction and the sensory input, then the residual errors between the two, are propagated bottom-up. Overlapping with some previous theories (Mumford, 1992; Barlow, 1994), the forward and backward connections are suggested to convey bottom-up *prediction errors* and the top-down predictions respectively.
>
> A key feature of the generative models is plasticity. The internal model is proposed to update its probabilistic history of past perceptions and their causes, i.e. *recognition density* or *priors*, to fine-tune itself iteratively with every prediction and increase the accuracy of future predictions. Both the predictions and prediction errors are relayed with *precision weighting*, i.e. an estimate of uncertainty, computed across all levels of the information processing cascade. The iterative updates of the internal model are influenced by the relative precision weights of the prediction and prediction error, where larger weights have greater impact on the distribution. The account therefore, puts forward hypotheses for the functional specialisation of connections in the anatomy of cortical hierarchies as well as dynamic process of prediction-to-perception. Any disruptions to the components of the model could result in domain-general impairments, for example over-reliance on the priors, under-reliance on sensory evidence, failure to detect errors, inability to capture probabilistic patterns and learn.
>
> In predictive coding, updating of the predictions is weighted by measures of certainty (or precision) of the predictions and the sensory information (Brown *et al.*, 2013; Palmer *et al.*, 2019), such that, in a novel environment sensory information is more likely to be weighted favourably, or when sensory information is diminished predictions are held with higher precision. This fine balance determines optimal motor control for initiating and inhibiting movement, and for learning or tuning motor schema. Although in predictive coding, dopamine is considered to be one of the key neurotransmitters mediating precision (Friston *et al.*, 2012; Friston *et al.*, 2014a), this theory is relevant to many non-dopaminergic disorders. Impairments at any level of the distributed hierarchical process for movement can result in a cascade of errors that leads to erroneous motor output that is difficult to modify. At lower levels, impairments lead to bradykinesia, apraxia, or alien limb, while dysfunction at higher levels can lead to apathy, impulsivity or disinhibition.

also worse at adapting to expected auditory stimuli (reduced auditory mismatch negativity responses - (Gaeta *et al.*, 1999; Pekkonen *et al.*, 2001; Laptinskaya *et al.*, 2018)). Even otherwise healthy APOE4 carriers (i.e. elevated risk of Alzheimer's disease) show impairments in detecting auditory targets using contextual information (Zimmermann *et al.*, 2019). Patients with amnestic and logopenic phenotypes of Alzheimer's disease are impaired in processing a melodic contour, which depends on working memory to predict the upcoming sounds (Golden *et al.*, 2017).

In the visual domain, hallucinations and illusions are commonly reported in patients with cortical Lewy body pathology. The perceptual content is often based on the immediate environment or autobiographical memories, with pareidolic experiences in ambiguous sceneries (Uchiyama *et al.*, 2012), or familiar people or pets (Barnes and David, 2001), even if known to have died. The hallucinations are visually complex and familiar, rather than simple visual percepts such as amorphous shapes and shadows (Collerton *et al.*, 2005; Mosimann *et al.*, 2006; Moran *et al.*, 2013). This is expected in the predictive coding framework, as a result of abnormal up-weighting of intermediate level priors and relative down-weighting of the visual sensory evidence (Friston, 2005b; Fletcher and Frith, 2009; Sterzer *et al.*, 2018; Corlett *et al.*, 2019). Several neuroanatomical sites have been implicated (Pezzoli *et al.*, 2017), with abnormal activity and connectivity between visual cortex, medial temporal and medial prefrontal areas (Stebbins *et al.*, 2004; Ramírez-Ruiz *et al.*, 2007; Perneczky *et al.*, 2008; Sanchez-Castaneda *et al.*, 2010; Peraza *et al.*, 2014; Heitz *et al.*, 2015; Shine *et al.*, 2015; Yao *et al.*, 2015; O'Callaghan *et al.*, 2017a). However, cholinergic insufficiency rather than atrophy is proposed to be the cause. Acetylcholine enhances sensory precision and strengthens bottom-up signalling (Moran *et al.*, 2013). The cholinergic loss in dementia with Lewy bodies would weaken the feed-forward prediction errors relative to the feedback information of predictions based on higher level priors (Collerton *et al.*, 2005; Diederich *et al.*, 2005; O'Callaghan *et al.*, 2017b). Indeed, patients who experience more visual hallucinations have more severe degeneration of their cholinergic pathways (Ballard *et al.*, 2000; Harding *et al.*, 2002; Halliday, 2005), and these symptoms are alleviated with cholinesterase inhibitors (Mori *et al.*, 2006).

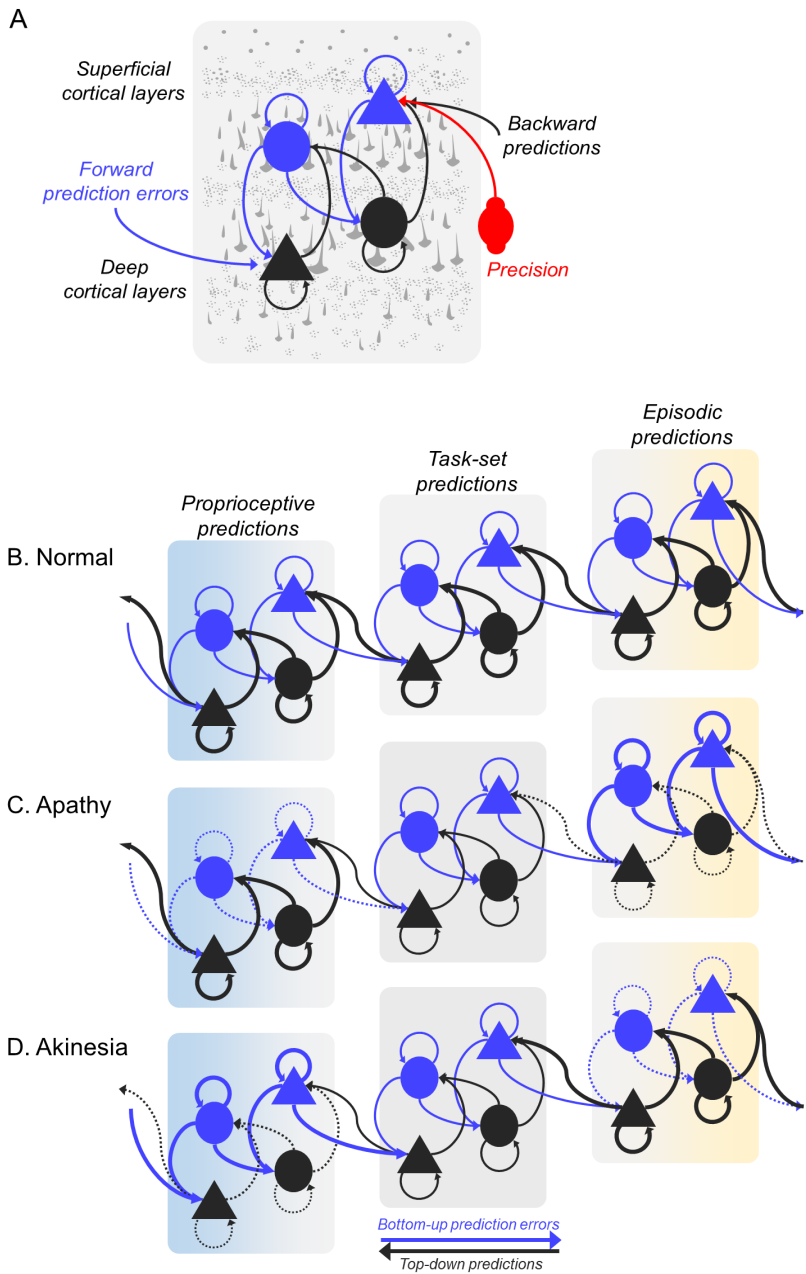

**Fig 1. Predictive coding mechanism within the hierarchical brain network**
**A.** Schematic illustration of the predictive coding mechanism at a single cortical layer within the hierarchy. Top-down predictions are conveyed via the backward connections (black arrows) from the state units (black nodes) in the deep cortical layers. The predictions are compared with conditional expectations at the lower level in the hierarchy by the error units in the superficial cortical layers (blue nodes) to produce prediction errors, which are then passed bottom-up (blue arrows) to the higher level to update the predictions in a Bayesian fashion. Triangles and circles represent pyramidal neurons and inhibitory interneurons respectively. Precision weighting (red) regulates the post-synaptic gain of the error units, via neuromodulation. B-D illustrates three layers of a hierarchical network of the motor system as an example of how the predictive coding mechanism gets impaired in neurodegenerative diseases, going from lower to higher cortical layers from left (light blue) to right (yellow). Each layer of the hierarchy makes predictions relayed in a top-down fashion. Higher layers of the network make episodic predictions that are multimodal, abstract and span across a longer timescale (e.g. *remembering that the London marathon is in two months*). Intermediate layers carry out task-set predictions that depend on the immediate context and last a shorter timescale (e.g. *expecting to see runners, cheering supporters and water stands on the venue*). Lower layers make fast-moving, proprioceptive predictions on the expected consequences of our actions (e.g. *expected position of the limbs and the body following each stride*). **B.** Normal cortical hierarchy of the motor system is characterized with optimal control where top-down predictions suppress the bottom-up prediction errors at each layer. **C.** In apathy, predictions at the higher levels fail to suppress prediction errors at the intermediate level, leading to stronger contextual prediction errors, and reduced goal-directed movement. **D.** In akinesia, predictions at the lower levels fail to suppress proprioceptive prediction errors, resulting in failure to move.

## Action and apathy

In the active inference framework, the prediction errors in sensory systems could be minimised either by updating future predictions or by changing the sensory inputs to match the predictions. Although the direct evidence for active inference is largely from motor control (Kilner, 2011), behavioural symptoms like apathy could arise from disruptions at higher levels (Hezemans *et al.*, 2020).

Evidence for active inference comes from 'sensorimotor attenuation': a transient down-weighting of the predicted sensory consequences of actions, observed in 98% of healthy adults (**Figure 1B**) (Wolpe *et al.*, 2016a). For example, when participants attempt to match a force applied to their hand by pressing a sensor with a finger, the force generated is typically greater than the force applied. This is suggested to facilitate movement, enhance perception, and provide a sense of agency (cf.(Wolpe and Rowe, 2014)). In healthy ageing there is greater reliance on the predictions and less on the sensorium (Wolpe *et al.*, 2016a). Whereas in neurodegenerative diseases, deficits in sensorimotor predictions (or their precision) results in an over-reliance on sensory evidence, causing a poverty of movement (Brown *et al.*, 2013; Wolpe *et al.*, 2016b; Wolpe *et al.*, 2018a). Deficits in sensorimotor predictions are linked to disease severity (Wolpe *et al.*, 2014; Wolpe *et al.*, 2018b), volumetric and white matter loss in pre-supplementary motor area (Halliday *et al.*, 2005; Wolpe and Rowe, 2014; Wolpe *et al.*, 2018a). Symptomatic therapies using peripheral vibration can improve motor symptoms in some patients (cf.Sweeney *et al.*, 2019 for review), by reducing the precision from sensory evidence and increasing the relative precision of the prediction (Macerollo *et al.*, 2018). The physiological correlate of sensorimotor attenuation is beta desynchronisation (Palmer *et al.*, 2016; Tan *et al.*, 2016; Palmer *et al.*, 2019) which is required for movement planning and initiation (Pfurtscheller and Lopes da Silva, 1999). In bradykinetic disorders, beta power is elevated (Schnitzler and Gross, 2005; Levy *et al.*, 2010; Bizovicar *et al.*, 2014; Moisello *et al.*, 2015). Dopaminergic treatment in Parkinson's disease can enhance beta desynchronisation, (Brown and Marsden, 1999; Levy *et al.*, 2010) and increase sensorimotor attenuation (Macerollo *et al.*, 2016; Wolpe *et al.*, 2018a).

Apathy, like bradykinesia, could be explained by deficits in the precision of the prediction, however the deficits occur within high levels of the hierarchy (**Figure 1C**): within a network involving anterior cingulate and prefrontal cortex with loss of connectivity to the striatum (Le Heron *et al.*, 2018; Nobis and Husain, 2018; Passamonti *et al.*, 2018). In active inference terms, when the precision of the prediction is low or there is low certainty in action-outcome mapping, the outcome is a lack of response (Friston *et al.*, 2010; Friston *et al.*, 2014a; Parr *et al.*, 2019). In healthy controls greater expression of apathy trait is associated with lower certainty on predictions about action outcomes (Hezemans *et al.*, 2020). In bradykinesia a lack of movement, for instance to switch on a light in a dark room, is due to overriding sensory evidence from proprioception that they are not moving, relative to the impaired precision on the predictions for movement. This results in limited sensorimotor attenuation, and thus an inability to initiate the action to switch on the light. In contrast, patients with apathy may not initiate the action to switch on the light because the sensory evidence, that the room is dark, overrides the weak predictions/precision of the internal prediction (to switch on the light).

Apathy may also be linked to deficits in how reward and cost are encoded in the prediction or how the value of the action weights precision (Hezemans *et al.*, 2020), suggesting that high cost or low value actions, and low certainty of action-outcome contingencies can result in absence of movement. In this light, dopamine is re-framed as a modulator of the precision of higher-order states, not just of the sensory evidence. As such, it modulates active inference by which complex behaviours are executed to resolve the prediction error between high-order predictions and intermediate feedforward evidence of the state. Devaluation of outcomes by dopamine depletion then reduces behaviour (Hezemans *et al.*, 2020). However, non-dopaminergic changes in dementia will lead to a similar change in precision weighting and result in apathy, including noradrenaline (Ruthirakuhan *et al.*, 2018), GABA and glutamate, which regulate the precision of feedforward and feedback information transfer in cortical hierarchies (Moran *et al.*, 2007; Moran *et al.*, 2015). The changes in GABA and glutamate in dementias (Murley and Rowe, 2018) may therefore contribute to apathy, in the presence of relatively normal dopaminergic function.

## Speech and language

Healthy language comprehension shows remarkable speed and resistance to noise, which is supported by predictive coding mechanisms at multiple levels of linguistic representation: phonological (Gagnepain *et al.*, 2012; Ettinger *et al.*, 2014; Monsalve *et al.*, 2018), semantic (DeLong *et al.*, 2005; Lau *et al.*, 2013; Lau and Nguyen, 2015; Maess *et al.*, 2016; Wang *et al.*, 2018; Klimovich-Gray *et al.*, 2019), syntactic (Fonteneau, 2013; Wlotko and Federmeier, 2015; Henderson *et al.*, 2016) and discourse context (Otten and Van Berkum, 2008). In neurodegenerative aphasias, many of the deficits of frontotemporal and temporo-parietal networks can be understood in terms of impairments of predictive coding.

Degeneration of frontal and perisylvian cortex leads to speech production deficits and agrammatism (Gorno-Tempini *et al.*, 2004; Hayes *et al.*, 2016; Henry *et al.*, 2016). This reduces the top-down control used to optimise perception and production of speech (Pickering and Garrod, 2007, 2013; Park *et al.*, 2015; Sohoglu and Davis, 2016). As a result, non-fluent aphasic patients show greater speech processing deficits and delays at the lexical level when speech is degraded (Utman *et al.*, 2001; Moineau *et al.*, 2005) or ambiguous (Hagoort, 1993; Swaab *et al.*, 1998; Grindrod and Baum, 2005). While damage to the temporo-parietal junction leads to repetition deficit (Baldo *et al.*, 2008; Buchsbaum *et al.*, 2011) arising from disrupted mapping between speech representations and proprioceptive articulatory predictions in the motor and inferior frontal cortices (Adams *et al.*, 2013; Parr *et al.*, 2018). Cope et al. (2017) showed that in the presence of intact temporal cortex, frontal neurodegeneration in non-fluent primary progressive aphasia (nvPPA) causes overly precise and inflexible contextual predictions, with reduced frontal-to-temporal directional connectivity (**Figure 2B-D**). This leads to delayed resolution of speech inputs by the temporal cortex, and impaired perception of degraded speech. However, the reliance on inflexible priors becomes a paradoxical advantage as noise increases. Accordingly, the patients' symptoms were relatively reduced in noisy environments and worse in quiet settings. Inflexible predictions similarly affect speech production in nvPPA. Whereas delayed auditory feedback in healthy controls reduces fluency and accuracy of speech (Lin *et al.*, 2015; Huang *et al.*, 2016), delayed feedback does not impair nvPPA fluency: this suggests a reliance on internal models of speech (with strong priors) and relative weakness of the precision of sensory representations (Hardy *et al.*, 2018).

Efficient reading requires greater top-down signalling from higher order language areas to disambiguate visually confusable words (Price and Devlin, 2011). While damage to the left medial occipito-temporal areas causes alexia and object agnosia with spared central language abilities and orthographic knowledge (Damasio and Damasio, 1983; Binder and Mohr, 1992), reading deficits are often more severe than object recognition deficits. Concurrently, lesions of inferior frontal cortex cause auditory agnosias and pure word deafness (Confavreux *et al.*, 1992; Otsuki *et al.*, 1998). Woodhead et al. (2013) showed that whole-word training to improve reading was associated with stronger feedback connectivity from the inferior frontal gyrus to the occipital areas, and bidirectional connectivity between ventral occipito-temporal and occipital areas. This suggests stronger top-down priors aid prediction of the words. Semantic processing is similarly dependent on top-down signalling, using contextual information and prior knowledge to predict forthcoming words (Kocagoncu *et al.*, 2017; Klimovich-Gray *et al.*, 2019; Lyu *et al.*, 2019). The N400 is an electrophysiological index of the prediction error, reflecting the degree of mismatch between semantic priors and sensory input (Kutas and Federmeier, 2011) (i.e. semantic prediction error). Semantic dementia impairs the differentiation of concepts that belong to the same semantic category, such as *giraffe* and *zebra* (i.e. taxonomic blurring).

They display preserved N400 to semantically unrelated words, but significantly reduced N400 to semantically related words (Hurley *et al.*, 2012), indicating the weakness of the semantic priors. Disambiguating meaningful objects (but not meaningless shapers) in difficult viewing conditions (Cumming *et al.*, 2006) is also impaired, suggesting domain-general deficit of top-down semantic control.

## Memory and learning

Statistical dependencies and regularities underpin our learned beliefs, and form the priors to understand future experience. Hippocampus is proposed to encode expectancies of future events based on the probabilistic consequences of the past events (Eichenbaum *et al.*, 1999; Strange *et al.*, 2005; Harrison *et al.*, 2006), its activity is modulated by the entropy, a measure of predictability, of the future events before they take place (Weiler *et al.*, 2010). A corollary of this, damage to the medial temporal lobe structures in many dementias, has severe implications for memory retrieval, episodic future thinking and probabilistic learning.

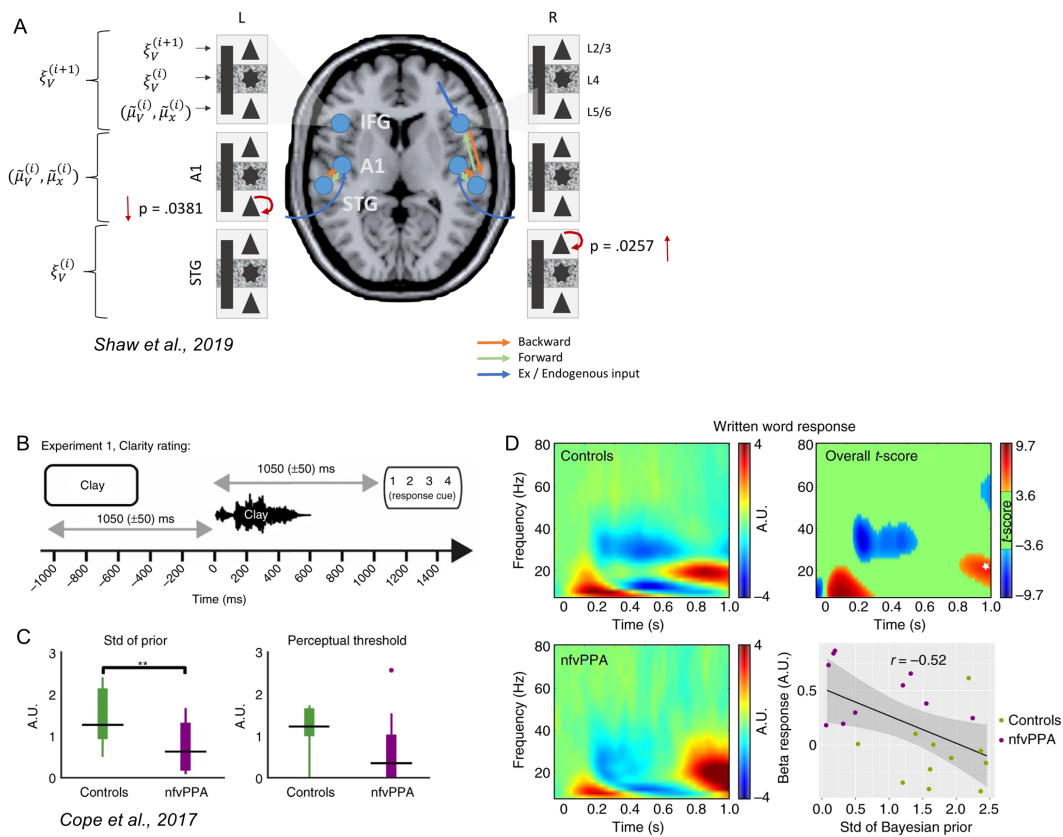

Fig 2. Neurophysiological changes associated with predictive coding impairments
**A.** Results of the cortical microcircuit dynamic causal modelling of the mismatch negativity responses in behavioural variant frontotemporal dementia patients compared to healthy controls. Figure displays local (intrinsic) decreases in self-modulation of the deep pyramidal cells in the primary auditory cortex (A1), and increases in self-modulation of the superficial pyramidal cells in the superior temporal gyrus, that underpins reduced mismatch responses and failure to establish sensory predictions. Reprinted from Shaw et al, 2019 with permission. B-D are reprinted from Cope et al, 2017 with permission. **B.** Illustration of the MEG paradigm. Participants were presented with a written word followed by a noise vocoded spoken word that either matched or mismatched with the written word. Participants rated the clarity of the spoken words. **C.** Derived parameters from the Bayesian data modelling showing differences in the standard deviation of the prior expectations (left panel) and perceptual thresholds between nvPPA and healthy controls. nvPPA patients had significantly more precise prior expectations than controls. A.U.: Arbitrary units. **D.** Induced time-frequency responses at the written word offset, and spoken word onset. The beta power was significantly higher in the nvPPA group after 800 ms. The beta power in this late peak negatively correlated with precision of the prior expectations.

Predictive coding mechanism underlies anticipating events both at micro and macro timescales. At the macro timescale, prospection refers to the ability to create mental simulations of future episodes, actions and expectations on their consequences (Gilbert and Wilson, 2007). Prospection is an integral part of decision making, and planning in novel scenarios. Future simulations are created by using prior episodes and knowledge as building blocks (Johnson and Sherman, 1990; Cohen and Eichenbaum, 1993), and they activate the same network involved in remembering the past: the prefrontal cortex and medial temporal lobe structures (Addis *et al.*, 2007; Schacter *et al.*, 2007). Patients with Alzheimer's disease, semantic dementia, and hippocampal damage show comparable impairments in past and future thinking (Hassabis and Maguire, 2007; Addis *et al.*, 2009; Andelman *et al.*, 2010; Gamboz *et al.*, 2010; Irish *et al.*, 2012; Irish *et al.*, 2015). Their imaginary episodes are fragmented and show significant reductions in richness and content. Similarly, selective insult to the prefrontal cortex (Klein *et al.*, 2002), in frontotemporal dementia (Irish *et al.*, 2013) and damage to the fronto-striatal pathways in Parkinson's disease (de Vito *et al.*, 2012) results in impoverished future thinking.

The ability to anticipate events at the micro timescale breaks down in many dementias. The weather prediction task captures probabilistic learning and is associated with activity and connectivity in fronto-striatal circuits (Knowlton *et al.*, 1994). Although patients with Alzheimer's can perform this task well (Klimkowicz-Mrowiec *et al.*, 2008), Parkinson's disease (Knowlton *et al.*, 1996; Shohamy *et al.*, 2004), Huntington's disease (Holl *et al.*, 2012), and frontotemporal dementia patients (Dalton *et al.*, 2012; Weickert *et al.*, 2013) have severely impaired performance. Short-term learning and plasticity are also identified by auditory mismatch paradigms. Mismatch responses to 'oddball events' index the prediction error that is fed-forward from primary to secondary auditory cortex then to frontal cortex, so as to update predictions that are in turn fed-backwards to temporal cortex (Garrido *et al.*, 2009). The amplitude of mismatch negativity is reduced in Alzheimer's disease, vascular dementia (Jiang *et al.*, 2017), Parkinson's disease (Brønnick *et al.*, 2010), and frontotemporal dementia (Hughes and Rowe, 2013), with impaired frontotemporal connectivity (Stam *et al.*, 2006; Hughes and Rowe, 2013; Beste *et al.*, 2017; Shaw *et al.*, 2019) (**Figure 2A**). Alzheimer's patients show impairments in the mismatch responses especially at longer inter-stimulus intervals (Pekkonen *et al.*, 1994; Gaeta *et al.*, 1999; Pekkonen *et al.*, 2001), in relation to reduced temporal activity (Ruzzoli *et al.*, 2016; Jiang *et al.*, 2017).

Memory and learning are dependent on cholinergic modulation of NMDA receptor plasticity (Miaskinov *et al.*, 2008), which modulates the precision of the prediction error (Moran *et al.*, 2013; Carbajal and Malmierca, 2018). Impaired mismatch response in Alzheimer's disease is partially explained by the degeneration of cholinergic projections, with relatively preserved top-down propagation of the priors (Ruzzoli *et al.*, 2016). Cholinergic agents partially restore the mismatch response in Alzheimer's disease (Engeland *et al.*, 2002), enhancing feed-forward signalling by precision of the sensory evidence weighting (Yu and Dayan, 2002; Moran *et al.*, 2013). Similarly, dopamine is suggested to encode prediction errors, and that successful feedback-based learning is dependent on intact dopaminergic modulation of bottom-up signals (Schultz *et al.*, 1997). Interestingly, when Parkinsons' patients are on medication, their learning is remediated on tasks that give positive feedback (Knowlton *et al.*, 1996; Frank *et al.*, 2004). These findings suggest that memory, sensory and reward-based learning could get impaired following dopaminergic, cholinergic imbalances as well as fronto-striatal damage in dementia.

## Risk taking and impulsivity

Disinhibited and impulsive behaviors are common to many dementias (Nombela *et al.*, 2014; Lansdall *et al.*, 2017; Borges *et al.*, 2019), describing the predisposition to act out of context, prematurely, or on the basis of little evidence (Dalley and Robbins, 2017). In the predictive coding framework, these behaviours may also be explained by impaired precision on the internal predictions at higher levels.

Early or fast responses, in tasks such as Go-NoGo or stop-signal RT, may be related to impaired elevated precision on prior beliefs (Limongi *et al.*, 2018). Drift diffusion models (Zhang *et al.*, 2016) demonstrate how patients with PSP can be both impulsive and bradykinetic in an oculomotor decision task: patients had slow drift rates, reflecting limited sensory attenuation (resulting in bradykinesia, as discussed above), but had a response bias towards the decision boundary – that is a high confidence in their prior 'to move'. Consequently, patients may be slow to move but nevertheless quick to reach a decision to move.

For more complex decisions, when selecting between different valued outcomes, impulsive decisions may be related to impairments in the prediction errors, as illustrated by classic gambling tasks. In gambling tasks, that pay high reward on 'risky decks', patients with frontotemporal dementia, Alzheimer's disease, and Parkinson's disease choose the risky decks more frequently than healthy controls (Delazer *et al.*, 2007; Sinz *et al.*, 2008) The risk-taking behaviour is associated with impairments in dopamine and noradrenaline signalling for processing prediction errors.

Dopamine and its receptor affinity is largely depleted in dorsal striatum and dorsolateral prefrontal cortex in early stages of Parkinson's disease (Agid *et al.*, 1993; Kaasinen *et al.*, 2003) and linked to impaired feedback processing (Brand *et al.*, 2004), specifically for modulating future decisions with the help of negative feedback (Frank *et al.*, 2004). These studies underline the crucial role of dopamine in prediction error signalling, as well as encoding reward-based outcome of decisions. The dopamine hypothesis may explain reflection impulsivity in Parkinson's disease (Averbeck *et al.*, 2013), and perhaps in part in PSP which has dopamine deficiency and significant atrophy of the midbrain (Bocchetta *et al.*, 2019).

Noradrenaline is also a key regulator of impulsivity. A focal noradrenergic area is the locus coeruleus, affected in many neurodegenerative diseases, including synucleinopathies, frontotemporal lobar degeneration and Alzheimer's disease (Betts *et al.*, 2019) and strongly associated with impulsivity (Passamonti *et al.*, 2018). The locus coeruleus is widely connected to motor control circuits, including the subthalamic nucleus, motor cortex, and pre-supplementary motor area (Hamani *et al.*, 2004; Bonnevie and Zaghloul, 2019). In predictive coding framework, activation in this structure is linked to learning about changing contingencies (for example in reversal learning) and is suggested to mediate the reciprocal subcortical-cortical circuit of updating predictions from prediction errors (Sales *et al.*, 2019). Depletion of tyrosine, a substrate required for dopamine synthesis, worsens performance on gambling tasks (Sevy *et al.*, 2006); whereas methylphenidate, a dopamine and noradrenergic agonist, ameliorates risk-taking in frontotemporal dementia patients (Rahman *et al.*, 2006). Similarly, atomoxetine, a selective noradrenaline reuptake inhibitor, reduces reflection impulsivity and risk taking in Parkinson's disease (Kehagia *et al.*, 2014; Rae *et al.*, 2016).

## Conclusion

In this Update, we reviewed recent clinical evidence with the main purpose of shifting our thinking from localist frameworks to disorders of cortical hierarchies, in understanding the makings of clinical phenomena. We have discussed the generalisability of the predictive coding principles to account for cognitive and perceptual impairments observed in many neurodegenerative diseases. Whilst acknowledging that predictive coding framework is not a panacea for explaining all clinical phenomena, we showed how a diverse range of neurocognitive deficits and aetiology could be described as mechanistic disruptions in a fine-tuned cortical system relying on maintaining a fragile balance. We discussed that there are multiple pathological routes leading to behavioural symptoms that appear similar on the surface but arise from different disruptions in the network. We then reviewed evidence on how the disruptions within hierarchical predictions could arise from changes in connectivity in relation to neurochemical imbalances that weight the importance (i.e. precision) of the predictions. Altogether, these studies demonstrate that cortical hierarchies could be thrown off balance in neurodegenerative diseases due to widespread atrophy, and changes in connectivity and neurochemistry, which could be explained within the predictive coding framework. Further frequency-resolved network and modelling at the microcircuit level of these mechanisms is necessary to further our understanding of clinical phenomena, and to develop better diagnostic and therapeutic tools.

# Funding

EK is funded by the Dementias Platform UK and Alzheimer's Research UK (RG94383/RG89702). JBR is supported by the Wellcome Trust (103838) and Medical Research Council (SUAG/004 RG91365). AKG is funded by the European Union's Horizon 2020 Research and Innovation Programme under the Marie Skłodowska-Curie grant (798971). LH is funded by the Wellcome Trust (103838).

# Competing interests

The authors report no competing interests.

# References

Adams RA, Shipp S, Friston KJ. Predictions not commands: active inference in the motor system. Brain Struct Funct 2013; 218(3): 611-43.

Addis DR, Sacchetti DC, Ally BA, Budson AE, Schacter DL. Episodic simulation of future events is impaired in mild Alzheimer's disease. Neuropsychologia 2009; 47(12): 2660-71.

Addis DR, Wong AT, Schacter DL. Remembering the past and imagining the future: common and distinct neural substrates during event construction and elaboration. Neuropsychologia 2007; 45(7): 1363-77.

Agid Y, Ruberg M, Javoy-Agid F, Hirsch E, Raisman-Vozari R, Vyas S, et al. Are dopaminergic neurons selectively vulnerable to Parkinson's disease? Adv Neurol 1993; 60: 148-64.

Andelman F, Hoofien D, Goldberg I, Aizenstein O, Neufeld MY. Bilateral hippocampal lesion and a selective impairment of the ability for mental time travel. Neurocase 2010; 16(5): 426-35.

Averbeck BB, Djamshidian A, O'Sullivan SS, Housden CR, Roiser JP, Lees AJ. Uncertainty about mapping future actions into rewards may underlie performance on multiple measures of impulsivity in behavioral addiction: evidence from Parkinson's disease. Behav Neurosci 2013; 127(2): 245-55.

Baldo JV, Klostermann EC, Dronkers NF. It's either a cook or a baker: patients with conduction aphasia get the gist but lose the trace. Brain Lang 2008; 105(2): 134-40.

Ballard C, Piggott M, Johnson M, Cairns N, Perry R, McKeith I, et al. Delusions associated with elevated muscarinic binding in dementia with Lewy bodies. Ann Neurol 2000; 48(6): 868-76.

Bar M. The proactive brain: using analogies and associations to generate predictions. Trends Cogn Sci 2007; 11(7): 280-9.

Barlow HB. What is the computational goal of the neocortex? In: Koch C, Davis JL, editors. Large-Scale Neuronal Theories of the Brain. Cambridge, MA: MIT Press; 1994.

Barnes J, David AS. Visual hallucinations in Parkinson's disease: a review and phenomenological survey. J Neurol Neurosurg Psychiatry 2001; 70(6): 727-33.

Beste C, Mückschel M, Rosales R, Domingo A, Lee L, Ng A, et al. Striosomal dysfunction affects behavioral adaptation but not impulsivity-Evidence from X-linked dystonia-parkinsonism. Mov Disord 2017; 32(4): 576-84.

Betts MJ, Kirilina E, Otaduy MCG, Ivanov D, Acosta-Cabronero J, Callaghan MF, et al. Locus coeruleus imaging as a biomarker for noradrenergic dysfunction in neurodegenerative diseases. Brain 2019; 142(9): 2558-71.

Binder JR, Mohr JP. The topography of callosal reading pathways. A case-control analysis. Brain 1992; 115 ( Pt 6): 1807-26.

Bizovicar N, Dreo J, Koritnik B, Zidar J. Decreased movement-related beta desynchronization and impaired post-movement beta rebound in amyotrophic lateral sclerosis. Clin Neurophysiol 2014; 125(8): 1689-99.

Bocchetta M, Iglesias JE, Chelban V, Jabbari E, Lamb R, Russell LL, et al. Automated Brainstem Segmentation Detects Differential Involvement in Atypical Parkinsonian Syndromes. J Mov Disord 2019.

Bonnevie T, Zaghloul KA. The Subthalamic Nucleus: Unravelling New Roles and Mechanisms in the Control of Action. Neuroscientist 2019; 25(1): 48-64.

Borges LG, Rademaker AW, Bigio EH, Mesulam MM, Weintraub S. Apathy and Disinhibition Related to Neuropathology in Amnestic Versus Behavioral Dementias. Am J Alzheimers Dis Other Demen 2019; 34(5): 337-43.

Brand M, Labudda K, Kalbe E, Hilker R, Emmans D, Fuchs G, et al. Decision-making impairments in patients with Parkinson's disease. Behav Neurol 2004; 15(3-4): 77-85.

Bregman AS. Auditory scene analysis: The perceptual organization of sound. Cambridge, Massachusetts: MIT Press; 1990.

Brown H, Adams RA, Parees I, Edwards M, Friston K. Active inference, sensory attenuation and illusions. Cogn Process 2013; 14(4): 411-27.

Brown P, Marsden CD. Bradykinesia and impairment of EEG desynchronization in Parkinson's disease. Mov Disord 1999; 14(3): 423-9.

Brønnick KS, Nordby H, Larsen JP, Aarsland D. Disturbance of automatic auditory change detection in dementia associated with Parkinson's disease: A mismatch negativity study. Neurobiol Aging 2010; 31(1): 104-13.

Buchsbaum BR, Baldo J, Okada K, Berman KF, Dronkers N, D'Esposito M, et al. Conduction aphasia, sensory-motor integration, and phonological short-term memory - an aggregate analysis of lesion and fMRI data. Brain Lang 2011; 119(3): 119-28.

Carbajal GV, Malmierca MS. The Neuronal Basis of Predictive Coding Along the Auditory Pathway: From the Subcortical Roots to Cortical Deviance Detection. Trends Hear 2018; 22: 2331216518784822.

Clark A. Whatever next? Predictive brains, situated agents, and the future of cognitive science. Behav Brain Sci 2013; 36(3): 181-204.

Cohen NJ, Eichenbaum H. Memory, Amnesia, and the Hippocampal System. Cambridge, MA: MIT Press; 1993.

Collerton D, Perry E, McKeith I. Why people see things that are not there: a novel Perception and Attention Deficit model for recurrent complex visual hallucinations. Behav Brain Sci 2005; 28(6): 737-57; discussion 57-94.

Confavreux C, Croisile B, Garassus P, Aimard G, Trillet M. Progressive amusia and aprosody. Arch Neurol 1992; 49(9): 971-6.

Cope TE, Sohoglu E, Sedley W, Patterson K, Jones PS, Wiggins J, *et al.* Evidence for causal top-down frontal contributions to predictive processes in speech perception. Nat Commun 2017; 8(1): 2154.

Corlett PR, Horga G, Fletcher PC, Alderson-Day B, Schmack K, Powers AR. Hallucinations and Strong Priors. Trends Cogn Sci 2019; 23(2): 114-27.

Cumming TB, Patterson K, Verfaellie M, Graham KS. One bird with two stones: Abnormal word length effects in pure alexia and semantic dementia. Cogn Neuropsychol 2006; 23(8): 1130-61.

Dalley JW, Robbins TW. Fractionating impulsivity: neuropsychiatric implications. Nat Rev Neurosci 2017; 18(3): 158-71.

Dalton MA, Weickert TW, Hodges JR, Piguet O, Hornberger M. Impaired acquisition rates of probabilistic associative learning in frontotemporal dementia is associated with fronto-striatal atrophy. Neuroimage Clin 2012; 2: 56-62.

Damasio AR, Damasio H. The anatomic basis of pure alexia. Neurology 1983; 33(12): 1573-83.

de Vito S, Gamboz N, Brandimonte MA, Barone P, Amboni M, Della Sala S. Future thinking in Parkinson's disease: An executive function? Neuropsychologia 2012; 50(7): 1494-501.

Delazer M, Sinz H, Zamarian L, Benke T. Decision-making with explicit and stable rules in mild Alzheimer's disease. Neuropsychologia 2007; 45(8): 1632-41.

DeLong KA, Urbach TP, Kutas M. Probabilistic word pre-activation during language comprehension inferred from electrical brain activity. Nat Neurosci 2005; 8(8): 1117-21.

Diederich NJ, Goetz CG, Stebbins GT. Repeated visual hallucinations in Parkinson's disease as disturbed external/internal perceptions: focused review and a new integrative model. Mov Disord 2005; 20(2): 130-40.

Dikker S, Pylkkänen L. Predicting language: MEG evidence for lexical preactivation. Brain Lang 2013; 127(1): 55-64.

Eichenbaum H, Dudchenko P, Wood E, Shapiro M, Tanila H. The hippocampus, memory, and place cells: is it spatial memory or a memory space? Neuron 1999; 23(2): 209-26.

Engeland C, Mahoney C, Mohr E, Ilivitsky V, Knott VJ. Acute nicotine effects on auditory sensory memory in tacrine-treated and nontreated patients with Alzheimer's disease: an event-related potential study. Pharmacol Biochem Behav 2002; 72(1-2): 457-64.

Ettinger A, Linzen T, Marantz A. The role of morphology in phoneme prediction: evidence from MEG. Brain Lang 2014; 129: 14-23.

Fletcher PC, Frith CD. Perceiving is believing: a Bayesian approach to explaining the positive symptoms of schizophrenia. Nat Rev Neurosci 2009; 10(1): 48-58.

Fonteneau E. Structural syntactic prediction measured with ELAN: evidence from ERPs. Neurosci Lett 2013; 534: 211-6.

Frank MJ, Seeberger LC, O'reilly RC. By carrot or by stick: cognitive reinforcement learning in parkinsonism. Science 2004; 306(5703): 1940-3.

Friston K, Schwartenbeck P, FitzGerald T, Moutoussis M, Behrens T, Dolan RJ. The anatomy of choice: dopamine and decision-making. Philos Trans R Soc Lond B Biol Sci 2014a; 369(1655).

Friston KJ. A theory of cortical responses. Philos Trans R Soc Lond B Biol Sci 2005a; 360(1456): 815-36.

Friston KJ. Hallucinations and perceptual inference. Behav Brain Sci 2005b; 28(6): 764-6.

Friston KJ, Daunizeau J, Kilner J, Kiebel SJ. Action and behavior: a free-energy formulation. Biol Cybern 2010; 102(3): 227-60.

Friston KJ, Shiner T, FitzGerald T, Galea JM, Adams R, Brown H, *et al.* Dopamine, affordance and active inference. PLoS Comput Biol 2012; 8(1): e1002327.

Friston KJ, Stephan KE, Montague R, Dolan RJ. Computational psychiatry: the brain as a phantastic organ. Lancet Psychiatry 2014b; 1(2): 148-58.

Gaeta H, Friedman D, Ritter W, Cheng J. Changes in sensitivity to stimulus deviance in Alzheimer's disease: an ERP perspective. Neuroreport 1999; 10(2): 281-7.

Gagnepain P, Henson RN, Davis MH. Temporal predictive codes for spoken words in auditory cortex. Curr Biol 2012; 22(7): 615-21.

Gamboz N, Brandimonte MA, De Vito S. The role of past in the simulation of autobiographical future episodes. Exp Psychol 2010; 57(6): 419-28.

Garrido MI, Kilner JM, Kiebel SJ, Friston KJ. Dynamic causal modeling of the response to frequency deviants. J Neurophysiol 2009; 101(5): 2620-31.

Gilbert DT, Wilson TD. Prospection: experiencing the future. Science 2007; 317(5843): 1351-4.

Golden HL, Clark CN, Nicholas JM, Cohen MH, Slattery CF, Paterson RW, *et al.* Music Perception in Dementia. J Alzheimers Dis 2017; 55(3): 933-49.

Golden HL, Nicholas JM, Yong KX, Downey LE, Schott JM, Mummery CJ, *et al.* Auditory spatial processing in Alzheimer's disease. Brain 2015; 138(Pt 1): 189-202.

Goll JC, Kim LG, Ridgway GR, Hailstone JC, Lehmann M, Buckley AH, *et al.* Impairments of auditory scene analysis in Alzheimer's disease. Brain 2012; 135(Pt 1): 190-200.

Gorno-Tempini ML, Dronkers NF, Rankin KP, Ogar JM, Phengrasamy L, Rosen HJ, *et al.* Cognition and anatomy in three variants of primary progressive aphasia. Ann Neurol 2004; 55(3): 335-46.

Griffiths TD, Warren JD. The planum temporale as a computational hub. Trends Neurosci 2002; 25(7): 348-53.

Grindrod CM, Baum SR. Hemispheric contributions to lexical ambiguity resolution in a discourse context: evidence from individuals with unilateral left and right hemisphere lesions. Brain Cogn 2005; 57(1): 70-83.
Hagoort P. Impairments of lexical-semantic processing in aphasia: evidence from the processing of lexical ambiguities. Brain Lang 1993; 45(2): 189-232.
Halliday G. Clarifying Lewy-body parkinsonism with visual hallucinations. Lancet Neurol 2005; 4(10): 588-9.
Halliday GM, Macdonald V, Henderson JM. A comparison of degeneration in motor thalamus and cortex between progressive supranuclear palsy and Parkinson's disease. Brain 2005; 128(Pt 10): 2272-80.
Hamani C, Saint-Cyr JA, Fraser J, Kaplitt M, Lozano AM. The subthalamic nucleus in the context of movement disorders. Brain 2004; 127(Pt 1): 4-20.
Harding AJ, Broe GA, Halliday GM. Visual hallucinations in Lewy body disease relate to Lewy bodies in the temporal lobe. Brain 2002; 125(Pt 2): 391-403.
Hardy CJD, Bond RL, Jaisin K, Marshall CR, Russell LL, Dick K*, et al.* Sensitivity of Speech Output to Delayed Auditory Feedback in Primary Progressive Aphasias. Front Neurol 2018; 9: 894.
Harrison LM, Duggins A, Friston KJ. Encoding uncertainty in the hippocampus. Neural Netw 2006; 19(5): 535-46.
Hassabis D, Maguire EA. Deconstructing episodic memory with construction. Trends Cogn Sci 2007; 11(7): 299-306.
Hayes RA, Dickey MW, Warren T. Looking for a Location: Dissociated Effects of Event-Related Plausibility and Verb-Argument Information on Predictive Processing in Aphasia. Am J Speech Lang Pathol 2016; 25(4S): S758-S75.
Heitz C, Noblet V, Cretin B, Philippi N, Kremer L, Stackfleth M*, et al.* Neural correlates of visual hallucinations in dementia with Lewy bodies. Alzheimers Res Ther 2015; 7(1): 6.
Henderson JM, Choi W, Lowder MW, Ferreira F. Language structure in the brain: A fixation-related fMRI study of syntactic surprisal in reading. Neuroimage 2016; 132: 293-300.
Henry ML, Wilson SM, Babiak MC, Mandelli ML, Beeson PM, Miller ZA*, et al.* Phonological Processing in Primary Progressive Aphasia. J Cogn Neurosci 2016; 28(2): 210-22.
Hezemans FH, Wolpe N, Rowe JB. Apathy is associated with reduced precision of prior beliefs about action outcomes. J Exp Psychol Gen 2020.
Hohwy J, Roepstorff A, Friston K. Predictive coding explains binocular rivalry: an epistemological review. Cognition 2008; 108(3): 687-701.
Holl AK, Wilkinson L, Tabrizi SJ, Painold A, Jahanshahi M. Probabilistic classification learning with corrective feedback is selectively impaired in early Huntington's disease--evidence for the role of the striatum in learning with feedback. Neuropsychologia 2012; 50(9): 2176-86.
Hosoya T, Baccus SA, Meister M. Dynamic predictive coding by the retina. Nature 2005; 436(7047): 71-7.
Huang X, Chen X, Yan N, Jones JA, Wang EQ, Chen L*, et al.* The impact of parkinson's disease on the cortical mechanisms that support auditory-motor integration for voice control. Hum Brain Mapp 2016; 37(12): 4248-61.
Hughes LE, Rowe JB. The impact of neurodegeneration on network connectivity: a study of change detection in frontotemporal dementia. J Cogn Neurosci 2013; 25(5): 802-13.
Hurley RS, Paller KA, Rogalski EJ, Mesulam MM. Neural mechanisms of object naming and word comprehension in primary progressive aphasia. J Neurosci 2012; 32(14): 4848-55.
Irish M, Addis DR, Hodges JR, Piguet O. Considering the role of semantic memory in episodic future thinking: evidence from semantic dementia. Brain 2012; 135(Pt 7): 2178-91.
Irish M, Halena S, Kamminga J, Tu S, Hornberger M, Hodges JR. Scene construction impairments in Alzheimer's disease - A unique role for the posterior cingulate cortex. Cortex 2015; 73: 10-23.
Irish M, Hodges JR, Piguet O. Episodic future thinking is impaired in the behavioural variant of frontotemporal dementia. Cortex 2013; 49(9): 2377-88.
Jiang S, Yan C, Qiao Z, Yao H, Qiu X, Yang X*, et al.* Mismatch negativity as a potential neurobiological marker of early-stage Alzheimer disease and vascular dementia. Neurosci Lett 2017; 647: 26-31.
Johnson MK, Sherman SJ. Constructing and reconstructing the past and the future in the present. In: Higgins ET, Sorrentino RM, editors. Handbook of motivation and cognition: Foundations of social behaviour. New York, NY: Guilford Press; 1990. p. 482-526.
Kaasinen V, Aalto S, NAgren K, Hietala J, Sonninen P, Rinne JO. Extrastriatal dopamine D(2) receptors in Parkinson's disease: a longitudinal study. J Neural Transm (Vienna) 2003; 110(6): 591-601.
Kehagia AA, Housden CR, Regenthal R, Barker RA, Muller U, Rowe J*, et al.* Targeting impulsivity in Parkinson's disease using atomoxetine. Brain 2014; 137(Pt 7): 1986-97.
Kilner JM. More than one pathway to action understanding. Trends Cogn Sci 2011; 15(8): 352-7.
Klein SB, Loftus J, Kihlstrom JF. Memory and temporal experience: the effects of episodic memory loss on an amnesic patient's ability to remember the past and imagine the future. Soc Cogn 2002; 20: 353-79.
Klimkowicz-Mrowiec A, Slowik A, Krzywoszanski L, Herzog-Krzywoszanska R, Szczudlik A. Severity of explicit memory impairment due to Alzheimer's disease improves effectiveness of implicit learning. J Neurol 2008; 255(4): 502-9.
Klimovich-Gray A, Tyler LK, Randall B, Kocagoncu E, Devereux B, Marslen-Wilson WD. Balancing Prediction and Sensory Input in Speech Comprehension: The Spatiotemporal Dynamics of Word Recognition in Context. J Neurosci 2019; 39(3): 519-27.
Knowlton BJ, Mangels JA, Squire LR. A neostriatal habit learning system in humans. Science 1996; 273(5280): 1399-402.
Knowlton BJ, Squire LR, Gluck MA. Probabilistic classification learning in amnesia. Learn Mem 1994; 1(2): 106-20.
Kocagoncu E, Clarke A, Devereux BJ, Tyler LK. Decoding the Cortical Dynamics of Sound-Meaning Mapping. J Neurosci 2017; 37(5): 1312-9.

Kumar S, Sedley W, Nourski KV, Kawasaki H, Oya H, Patterson RD, et al. Predictive coding and pitch processing in the auditory cortex. J Cogn Neurosci 2011; 23(10): 3084-94.
Kutas M, Federmeier KD. Thirty years and counting: finding meaning in the N400 component of the event-related brain potential (ERP). Annu Rev Psychol 2011; 62: 621-47.
Lansdall CJ, Coyle-Gilchrist ITS, Jones PS, Vazquez Rodriguez P, Wilcox A, Wehmann E, et al. Apathy and impulsivity in frontotemporal lobar degeneration syndromes. Brain 2017; 140(6): 1792-807.
Laptinskaya D, Thurm F, Küster OC, Fissler P, Schlee W, Kolassa S, et al. Auditory Memory Decay as Reflected by a New Mismatch Negativity Score Is Associated with Episodic Memory in Older Adults at Risk of Dementia. Front Aging Neurosci 2018; 10: 5.
Lau EF, Holcomb PJ, Kuperberg GR. Dissociating N400 effects of prediction from association in single-word contexts. J Cogn Neurosci 2013; 25(3): 484-502.
Lau EF, Nguyen E. The role of temporal predictability in semantic expectation: An MEG investigation. Cortex 2015; 68: 8-19.
Lawson RP, Rees G, Friston KJ. An aberrant precision account of autism. Front Hum Neurosci 2014; 8: 302.
Le Heron C, Apps MAJ, Husain M. The anatomy of apathy: A neurocognitive framework for amotivated behaviour. Neuropsychologia 2018; 118(Pt B): 54-67.
Levy R, Lozano AM, Lang AE, Dostrovsky JO. Event-related desynchronization of motor cortical oscillations in patients with multiple system atrophy. Exp Brain Res 2010; 206(1): 1-13.
Lewis AG, Bastiaansen M. A predictive coding framework for rapid neural dynamics during sentence-level language comprehension. Cortex 2015; 68: 155-68.
Lewis AG, Wang L, Bastiaansen M. Fast oscillatory dynamics during language comprehension: Unification versus maintenance and prediction? Brain Lang 2015; 148: 51-63.
Limongi R, Bohaterewicz B, Nowicka M, Plewka A, Friston KJ. Knowing when to stop: Aberrant precision and evidence accumulation in schizophrenia. Schizophr Res 2018; 197: 386-91.
Lin IF, Mochida T, Asada K, Ayaya S, Kumagaya S, Kato M. Atypical delayed auditory feedback effect and Lombard effect on speech production in high-functioning adults with autism spectrum disorder. Front Hum Neurosci 2015; 9: 510.
Lyu B, Choi HS, Marslen-Wilson WD, Clarke A, Randall B, Tyler LK. Neural dynamics of semantic composition. Proc Natl Acad Sci U S A 2019; 116(42): 21318-27.
Macerollo A, Chen JC, Korlipara P, Foltynie T, Rothwell J, Edwards MJ, et al. Dopaminergic treatment modulates sensory attenuation at the onset of the movement in Parkinson's disease: A test of a new framework for bradykinesia. Mov Disord 2016; 31(1): 143-6.
Macerollo A, Palmer C, Foltynie T, Korlipara P, Limousin P, Edwards M, et al. High-frequency peripheral vibration decreases completion time on a number of motor tasks. Eur J Neurosci 2018; 48(2): 1789-802.
Maess B, Mamashli F, Obleser J, Helle L, Friederici AD. Prediction Signatures in the Brain: Semantic Pre-Activation during Language Comprehension. Front Hum Neurosci 2016; 10: 591.
Miaskinov AA, Chen JC, Weinberger NM. Specific auditory memory induced by nucleus basalis stimulation depends on intrinsic acetylcholine. Neurobiol Learn Mem 2008; 90(2): 443-54.
Moineau S, Dronkers NF, Bates E. Exploring the processing continuum of single-word comprehension in aphasia. J Speech Lang Hear Res 2005; 48(4): 884-96.
Moisello C, Blanco D, Lin J, Panday P, Kelly SP, Quartarone A, et al. Practice changes beta power at rest and its modulation during movement in healthy subjects but not in patients with Parkinson's disease. Brain Behav 2015; 5(10): e00374.
Monsalve IF, Bourguignon M, Molinaro N. Theta oscillations mediate pre-activation of highly expected word initial phonemes. Sci Rep 2018; 8(1): 9503.
Moran RJ, Campo P, Symmonds M, Stephan KE, Dolan RJ, Friston KJ. Free energy, precision and learning: the role of cholinergic neuromodulation. J Neurosci 2013; 33(19): 8227-36.
Moran RJ, Jones MW, Blockeel AJ, Adams RA, Stephan KE, Friston KJ. Losing control under ketamine: suppressed cortico-hippocampal drive following acute ketamine in rats. Neuropsychopharmacology 2015; 40(2): 268-77.
Moran RJ, Kiebel SJ, Stephan KE, Reilly RB, Daunizeau J, Friston KJ. A neural mass model of spectral responses in electrophysiology. Neuroimage 2007; 37(3): 706-20.
Mori S, Mori E, Iseki E, Kosaka K. Efficacy and safety of donepezil in patients with dementia with Lewy bodies: preliminary findings from an open-label study. Psychiatry Clin Neurosci 2006; 60(2): 190-5.
Mosimann UP, Rowan EN, Partington CE, Collerton D, Littlewood E, O'Brien JT, et al. Characteristics of visual hallucinations in Parkinson disease dementia and dementia with lewy bodies. Am J Geriatr Psychiatry 2006; 14(2): 153-60.
Mumford D. On the computational architecture of the neocortex. II. The role of cortico-cortical loops. Biol Cybern 1992; 66(3): 241-51.
Murley AG, Rowe JB. Neurotransmitter deficits from frontotemporal lobar degeneration. Brain 2018; 141(5): 1263-85.
Nobis L, Husain M. Apathy in Alzheimer's disease. Curr Opin Behav Sci 2018; 22: 7-13.
Nombela C, Rittman T, Robbins TW, Rowe JB. Multiple modes of impulsivity in Parkinson's disease. PLoS One 2014; 9(1): e85747.
O'Callaghan C, Hall JM, Tomassini A, Muller AJ, Walpola IC, Moustafa AA, et al. Visual Hallucinations Are Characterized by Impaired Sensory Evidence Accumulation: Insights From Hierarchical Drift Diffusion Modeling in Parkinson's Disease. Biol Psychiatry Cogn Neurosci Neuroimaging 2017a; 2(8): 680-8.
O'Callaghan C, Kveraga K, Shine JM, Adams RB, Bar M. Predictions penetrate perception: Converging insights from brain, behaviour and disorder. Conscious Cogn 2017b; 47: 63-74.
O'Doherty JP, Buchanan TW, Seymour B, Dolan RJ. Predictive neural coding of reward preference involves dissociable responses in human ventral midbrain and ventral striatum. Neuron 2006; 49(1): 157-66.
Otsuki M, Soma Y, Sato M, Homma A, Tsuji S. Slowly progressive pure word deafness. Eur Neurol 1998; 39(3): 135-40.

Otten M, Van Berkum JJ. Discourse-based word anticipation during language processing: Prediction of priming? Discourse Process 2008; 45(6): 464-96.
Palmer C, Zapparoli L, Kilner JM. A New Framework to Explain Sensorimotor Beta Oscillations. Trends Cogn Sci 2016; 20(5): 321-3.
Palmer CE, Auksztulewicz R, Ondobaka S, Kilner JM. Sensorimotor beta power reflects the precision-weighting afforded to sensory prediction errors. Neuroimage 2019; 200: 59-71.
Park H, Ince RA, Schyns PG, Thut G, Gross J. Frontal top-down signals increase coupling of auditory low-frequency oscillations to continuous speech in human listeners. Curr Biol 2015; 25(12): 1649-53.
Parr T, Rees G, Friston KJ. Computational Neuropsychology and Bayesian Inference. Front Hum Neurosci 2018; 12: 61.
Parr T, Rikhye RV, Halassa MM, Friston KJ. Prefrontal Computation as Active Inference. Cereb Cortex 2019.
Passamonti L, Lansdall CJ, Rowe JB. The neuroanatomical and neurochemical basis of apathy and impulsivity in frontotemporal lobar degeneration. Curr Opin Behav Sci 2018; 22: 14-20.
Pekkonen E, Hirvonen J, Jääskeläinen IP, Kaakkola S, Huttunen J. Auditory sensory memory and the cholinergic system: implications for Alzheimer's disease. Neuroimage 2001; 14(2): 376-82.
Pekkonen E, Jousmäki V, Könönen M, Reinikainen K, Partanen J. Auditory sensory memory impairment in Alzheimer's disease: an event-related potential study. Neuroreport 1994; 5(18): 2537-40.
Pellicano E, Burr D. When the world becomes 'too real': a Bayesian explanation of autistic perception. Trends Cogn Sci 2012; 16(10): 504-10.
Peraza LR, Kaiser M, Firbank M, Graziadio S, Bonanni L, Onofrj M*, et al*. fMRI resting state networks and their association with cognitive fluctuations in dementia with Lewy bodies. Neuroimage Clin 2014; 4: 558-65.
Perneczky R, Drzezga A, Boecker H, Förstl H, Kurz A, Häussermann P. Cerebral metabolic dysfunction in patients with dementia with Lewy bodies and visual hallucinations. Dement Geriatr Cogn Disord 2008; 25(6): 531-8.
Pezzoli S, Cagnin A, Bandmann O, Venneri A. Structural and Functional Neuroimaging of Visual Hallucinations in Lewy Body Disease: A Systematic Literature Review. Brain Sci 2017; 7(7).
Pfurtscheller G, Lopes da Silva FH. Event-related EEG/MEG synchronization and desynchronization: basic principles. Clin Neurophysiol 1999; 110(11): 1842-57.
Pickering MJ, Garrod S. Do people use language production to make predictions during comprehension? Trends Cogn Sci 2007; 11(3): 105-10.
Pickering MJ, Garrod S. How tightly are production and comprehension interwoven? Front Psychol 2013; 4: 238.
Price CJ, Devlin JT. The interactive account of ventral occipitotemporal contributions to reading. Trends Cogn Sci 2011; 15(6): 246-53.
Rae CL, Nombela C, Rodriguez PV, Ye Z, Hughes LE, Jones PS*, et al.* Atomoxetine restores the response inhibition network in Parkinson's disease. Brain 2016; 139(Pt 8): 2235-48.
Rahman S, Robbins TW, Hodges JR, Mehta MA, Nestor PJ, Clark L*, et al.* Methylphenidate ('Ritalin') can ameliorate abnormal risk-taking behavior in the frontal variant of frontotemporal dementia. Neuropsychopharmacology 2006; 31(3): 651-8.
Ramnani N, Miall RC. A system in the human brain for predicting the actions of others. Nat Neurosci 2004; 7(1): 85-90.
Ramírez-Ruiz B, Martí MJ, Tolosa E, Giménez M, Bargalló N, Valldeoriola F*, et al.* Cerebral atrophy in Parkinson's disease patients with visual hallucinations. Eur J Neurol 2007; 14(7): 750-6.
Rao RP, Ballard DH. Predictive coding in the visual cortex: a functional interpretation of some extra-classical receptive-field effects. Nat Neurosci 1999; 2(1): 79-87.
Ruthirakuhan MT, Herrmann N, Abraham EH, Chan S, Lanctôt KL. Pharmacological interventions for apathy in Alzheimer's disease. Cochrane Database Syst Rev 2018; 5: CD012197.
Ruzzoli M, Pirulli C, Mazza V, Miniussi C, Brignani D. The mismatch negativity as an index of cognitive decline for the early detection of Alzheimer's disease. Sci Rep 2016; 6: 33167.
Sales AC, Friston KJ, Jones MW, Pickering AE, Moran RJ. Locus Coeruleus tracking of prediction errors optimises cognitive flexibility: An Active Inference model. PLoS Comput Biol 2019; 15(1): e1006267.
Sanchez-Castaneda C, Rene R, Ramirez-Ruiz B, Campdelacreu J, Gascon J, Falcon C*, et al.* Frontal and associative visual areas related to visual hallucinations in dementia with Lewy bodies and Parkinson's disease with dementia. Mov Disord 2010; 25(5): 615-22.
Schacter DL, Addis DR, Buckner RL. Remembering the past to imagine the future: the prospective brain. Nat Rev Neurosci 2007; 8(9): 657-61.
Schnitzler A, Gross J. Normal and pathological oscillatory communication in the brain. Nat Rev Neurosci 2005; 6(4): 285-96.
Schultz W, Dayan P, Montague PR. A neural substrate of prediction and reward. Science 1997; 275(5306): 1593-9.
Sevy S, Hassoun Y, Bechara A, Yechiam E, Napolitano B, Burdick K*, et al.* Emotion-based decision-making in healthy subjects: short-term effects of reducing dopamine levels. Psychopharmacology (Berl) 2006; 188(2): 228-35.
Shaw AD, Hughes LE, Moran R, Coyle-Gilchrist I, Rittman T, Rowe JB. In Vivo Assay of Cortical Microcircuitry in Frontotemporal Dementia: A Platform for Experimental Medicine Studies. Cereb Cortex 2019.
Shine JM, Muller AJ, O'Callaghan C, Hornberger M, Halliday GM, Lewis SJ. Abnormal connectivity between the default mode and the visual system underlies the manifestation of visual hallucinations in Parkinson's disease: a task-based fMRI study. NPJ Parkinsons Dis 2015; 1: 15003.
Shohamy D, Myers CE, Grossman S, Sage J, Gluck MA, Poldrack RA. Cortico-striatal contributions to feedback-based learning: converging data from neuroimaging and neuropsychology. Brain 2004; 127(Pt 4): 851-9.
Sinz H, Zamarian L, Benke T, Wenning GK, Delazer M. Impact of ambiguity and risk on decision making in mild Alzheimer's disease. Neuropsychologia 2008; 46(7): 2043-55.
Sohoglu E, Davis MH. Perceptual learning of degraded speech by minimizing prediction error. Proc Natl Acad Sci U S A 2016; 113(12): E1747-56.

Stam CJ, Jones BF, Manshanden I, van Cappellen van Walsum AM, Montez T, Verbunt JP, *et al.* Magnetoencephalographic evaluation of resting-state functional connectivity in Alzheimer's disease. Neuroimage 2006; 32(3): 1335-44.
Stebbins GT, Goetz CG, Carrillo MC, Bangen KJ, Turner DA, Glover GH, *et al.* Altered cortical visual processing in PD with hallucinations: an fMRI study. Neurology 2004; 63(8): 1409-16.
Sterzer P, Adams RA, Fletcher P, Frith C, Lawrie SM, Muckli L, *et al.* The Predictive Coding Account of Psychosis. Biol Psychiatry 2018; 84(9): 634-43.
Strange BA, Duggins A, Penny W, Dolan RJ, Friston KJ. Information theory, novelty and hippocampal responses: unpredicted or unpredictable? Neural Netw 2005; 18(3): 225-30.
Summerfield C, de Lange FP. Expectation in perceptual decision making: neural and computational mechanisms. Nat Rev Neurosci 2014; 15(11): 745-56.
Swaab TY, Brown C, Hagoort P. Understanding ambiguous words in sentence contexts: electrophysiological evidence for delayed contextual selection in Broca's aphasia. Neuropsychologia 1998; 36(8): 737-61.
Sweeney D, Quinlan LR, Browne P, Richardson M, Meskell P, G OL. A Technological Review of Wearable Cueing Devices Addressing Freezing of Gait in Parkinson's Disease. Sensors (Basel) 2019; 19(6).
Tan H, Wade C, Brown P. Post-Movement Beta Activity in Sensorimotor Cortex Indexes Confidence in the Estimations from Internal Models. J Neurosci 2016; 36(5): 1516-28.
Uchiyama M, Nishio Y, Yokoi K, Hirayama K, Imamura T, Shimomura T, *et al.* Pareidolias: complex visual illusions in dementia with Lewy bodies. Brain 2012; 135(Pt 8): 2458-69.
Utman JA, Blumstein SE, Sullivan K. Mapping from sound to meaning: reduced lexical activation in Broca's aphasics. Brain Lang 2001; 79(3): 444-72.
Vuust P, Ostergaard L, Pallesen KJ, Bailey C, Roepstorff A. Predictive coding of music--brain responses to rhythmic incongruity. Cortex 2009; 45(1): 80-92.
Vuust P, Witek MA. Rhythmic complexity and predictive coding: a novel approach to modeling rhythm and meter perception in music. Front Psychol 2014; 5: 1111.
Wang L, Kuperberg G, Jensen O. Specific lexico-semantic predictions are associated with unique spatial and temporal patterns of neural activity. Elife 2018; 7.
Weickert TW, Leslie F, Rushby JA, Hodges JR, Hornberger M. Probabilistic association learning in frontotemporal dementia and schizophrenia. Cortex 2013; 49(1): 101-6.
Weiler JA, Suchan B, Daum I. Foreseeing the future: occurrence probability of imagined future events modulates hippocampal activation. Hippocampus 2010; 20(6): 685-90.
Wicha NY, Moreno EM, Kutas M. Anticipating words and their gender: an event-related brain potential study of semantic integration, gender expectancy, and gender agreement in Spanish sentence reading. J Cogn Neurosci 2004; 16(7): 1272-88.
Wlotko EW, Federmeier KD. Time for prediction? The effect of presentation rate on predictive sentence comprehension during word-by-word reading. Cortex 2015; 68: 20-32.
Wolpe N, Ingram JN, Tsvetanov KA, Geerligs L, Kievit RA, Henson RN, *et al.* Ageing increases reliance on sensorimotor prediction through structural and functional differences in frontostriatal circuits. Nat Commun 2016a; 7: 13034.
Wolpe N, Ingram JN, Tsvetanov KA, Geerligs L, Kievit RA, Henson RN, *et al.* Ageing increases reliance on sensorimotor prediction through structural and functional differences in frontostriatal circuits. Nat Commun 2016b; 7: 13034.
Wolpe N, Moore JW, Rae CL, Rittman T, Altena E, Haggard P, *et al.* The medial frontal-prefrontal network for altered awareness and control of action in corticobasal syndrome. Brain 2014; 137(Pt 1): 208-20.
Wolpe N, Rowe JB. Beyond the "urge to move": objective measures for the study of agency in the post-Libet era. Front Hum Neurosci 2014; 8: 450.
Wolpe N, Zhang J, Nombela C, Ingram JN, Wolpert DM, Cam CAN, *et al.* Sensory attenuation in Parkinson's disease is related to disease severity and dopamine dose. Sci Rep 2018a; 8(1): 15643.
Wolpe N, Zhang J, Nombela C, Ingram JN, Wolpert DM, Rowe JB, *et al.* Sensory attenuation in Parkinson's disease is related to disease severity and dopamine dose. Sci Rep 2018b; 8(1): 15643.
Woodhead ZV, Penny W, Barnes GR, Crewes H, Wise RJ, Price CJ, *et al.* Reading therapy strengthens top-down connectivity in patients with pure alexia. Brain 2013; 136(Pt 8): 2579-91.
Yao N, Pang S, Cheung C, Chang RS, Lau KK, Suckling J, *et al.* Resting activity in visual and corticostriatal pathways in Parkinson's disease with hallucinations. Parkinsonism Relat Disord 2015; 21(2): 131-7.
Yu AJ, Dayan P. Acetylcholine in cortical inference. Neural Netw 2002; 15(4-6): 719-30.
Zhang J, Rittman T, Nombela C, Fois A, Coyle-Gilchrist I, Barker RA, *et al.* Different decision deficits impair response inhibition in progressive supranuclear palsy and Parkinson's disease. Brain 2016; 139(Pt 1): 161-73.
Zimmermann J, Alain C, Butler C. Impaired memory-guided attention in asymptomatic APOE4 carriers. Sci Rep 2019; 9(1): 8138.